\documentclass[a4paper, 11pt]{article}

\usepackage{amsmath, amssymb}
\usepackage[a4paper,right=3cm,left=3cm,top=2.5cm,bottom=2.5cm]{geometry}
\usepackage{tikz}
\usepackage[linesnumbered,algoruled,boxed,lined]{algorithm2e}
\usepackage{xcolor}
\usepackage{natbib}
\usepackage{amsthm}

\theoremstyle{definition}
\newtheorem{example}{Example}
\newtheorem{definition}{Definition}
\newtheorem{proposition}{Proposition}
\newtheorem{theorem}{Theorem}

\newtheorem{remark}{Remark}

\begin{document}
	
	\title{The numerical statistical fan for noisy experimental designs}                       
	
	\author{Arkadius Kalka and Sonja Kuhnt}           
	\maketitle
	
	\begin{quotation}
		\noindent {\it Abstract:}
	Identifiability of polynomial models is a key requirement for multiple regression. We consider an analogue of the so-called statistical fan, the set of all maximal identifiable hierarchical models, for cases of noisy experimental designs or measured covariate vectors with a given tolerance vector. 
 This gives rise to the definition of the numerical statistical fan. It includes all maximal hierarchical models that avoid approximate linear dependence of the design vectors. We develop an algorithm to compute the  numerical statistical fan using recent results on the computation of all border bases of a design ideal from the field of algebra. 
The ideas are applied to data from a thermal spraying process. It turns out that the numerical statistical fan is effectively computable and much smaller than the respective statistical fan.  
The  gained enhanced knowledge of the
space of all stable identifiable  hierarchical models enables  improved model selection procedures.

\vspace{9pt}
\noindent {\it Key words and phrases:}
	Algebraic statistics, identifiable regression models, hierarchical  models, noisy experimental design,  statistical fan.
\par
\end{quotation}\par

\section{Introduction}
\label{Intro}

Model selection procedures play an integral part in regression analysis, with 
hierarchical polynomial models often being the maximal models considered.
For many experimental designs like full factorial designs or central composite designs,  the set of  identifiable hierarchical maximal models 
is well-known and usually quite small as these designs exhibit many symmetries. However, we might not have an experimental setup with a well-established experimental design. Then, the set of all identifiable maximal hierarchical models is given by the statistical fan from the field of algebraic statistics \citep{PRW00}. A selection and search algorithm in the set of hierarchical models is suggested in \cite{BGW03}. If we are  modeling some response $Y$ from a design $D_X$ given by a set of input data points $x_1,\ldots,x_n$, then - for generic input data -  the statistical fan is hard to enumerate, because its size grows (sub)-exponentially with
the number of data points. 
If the input data points are observed or measured, they  come with errors. Given a tolerance vector that bounds these errors, we may define sets of points of limited precision or noisy designs. This gives rise to the question which hierarchical models are identifiable for such empirical designs. Our goal is to formalize and exploit the notion of a  numerical statistical fans, which includes all maximal hierarchical models that avoid any kind of ``numerical aliasing'' between the model terms, i.e. any approximate linear dependence of the design vectors.

A subset of a  numerical statistical fan, the numerical algebraic fan,  has been previously introduced in \cite{RKR16}.
It is the numerical analogue of the algebraic fan which contains only hierarchical models that can 
be obtained by Gr{\"o}bner basis techniques and which is usually only a small part of the desired statistical fan \citep{Mar07}. 

In this contribution, we derive a recursive algorithm that effectively computes the numerical statistical fan, if the norm of the tolerance is not too small. Our algorithm is a modification of an algorithm proposed in \cite{HKP19} that computes all border bases (of a design ideal) and their order ideals.
Actually, the first algorithm that allows one to compute all 
border bases is given by \cite{BP16}.
 They provided a polyhedral characterisation of identifiable order ideals which are in one-to-one correspondence to integral points of the so-called order ideal polytope.

We apply our methods to real data coming from thermal spraying. There we have a  design $D_Y$ that is itself the response to a well chosen experimental design $D_X$. 
Section \ref{Prelim} 
introduces the needed background from algebraic statistics like order ideals, the statistical fan, Gr{\"o}bner bases and the
algebraic fan. Section \ref{NumFan} 
defines empirical designs, the notion of numerical linear dependence, stable
order ideals and the numerical statistical fan of a noisy design. There we also describe the recursive algorithm
to compute the numerical statistical fan. Section \ref{SprayData} 
deals with an application to thermal spraying data.
We compute the numerical statistical fan, its size distribution of stable order ideals and compare it to the statistical fan. Section \ref{sec:OutlookAndDiscussion} 
contains discussion and outlook.

\section{Background from algebraic statistics}
\label{Prelim}

In this section we revisit basic notions of algebraic statistics. In particular, Section \ref{Stfan} reveals  design ideals, hierarchical models, design matrices and the statistical fan.
Section \ref{Algfan} deals with Gr{\"o}bner bases and the algebraic fan.

\subsection{Hierarchical models and the statistical fan}\label{Stfan}

A typical situation in applications of statistical design of experiments is that $d$ controllable input factors $X=(X_1, \ldots ,X_d)^t$ influence a response $Y$. We run an experimental design $D$ with settings for $X$, observe response values $\{y(x)\}_{x\in D}$, and fit a linear regression model. 
In algebraic statistics a design $D$ is viewed as the common zero set of polynomials in a so-called design ideal.
\begin{definition}
	Let $R=K[X_1, \ldots,X_d]$ be the multivariate polynomial ring in $d$ variables $X_1, \ldots ,X_d$ over the field $K$ (with $K=\mathbb{Q}$ or $\mathbb{R}$). A \emph{design} $D=\{p_1, \ldots , p_n\}$, $n \in \mathbb{N}$, is a finite set of points in $K^d$.
	The \emph{design ideal} $I(D)$ is the set of all polynomials in $R$ that vanish at the design points.
\end{definition}
The set of all terms/monomials in $R$ is denoted by $\mathcal{T}=\{X_1^{\alpha_1}\cdots X_d^{\alpha_d}\mid \alpha_i \ge 0, \quad i=1,\ldots ,d\}$. This set is in one-to-one correspondence to $\mathbb{N}^d$ via some discrete logarithm map
\[ \log : \quad X^{\alpha }=X_1^{\alpha_1}\cdots X_d^{\alpha_d} \mapsto \alpha=(\alpha _1, \ldots ,\alpha _d). \]
In general we are fitting polynomial models of the form $\sum_{\alpha\in T} \beta_{\alpha}X^{\alpha}$
with $T\subseteq \mathbb{N}^d$. Thus a polynomial model can be viewed as a finite set of terms and is completely described by the 
finite subset $T$ of $\mathbb{N}^d$. Of particular importance are hierarchical polynomial models, i.e. for any higher order term, the model also contains all of the lower order terms that compose it, e.g. with an interaction $X_1X_2$ also $1$, $X_1$ and $X_2$ are contained in the model.
\begin{definition}
	A \emph{hierarchical model} or \emph{order ideal}  is a finite subset $\mathcal{O}$ of $\mathcal{T}$ that
	is closed under divisibility, i.e. $t\in \mathcal{O}$ implies $t' \in \mathcal{O}$ for all $t'\mid t$.
\end{definition}
Note that $X^{\alpha}$ divides $X^{\gamma}$ if and only if $X^{\alpha_i}_i$ divides $X^{\gamma_i}_i$ for all $1\le i \le d$. Hence divisibility of terms is mapped to the natural partial order on $\mathbb{N}^d$ via the above mentioned discrete logarithm.
The subset $\log \mathcal{O} \subseteq \mathbb{N}^d$ corresponding to an order ideal $\mathcal{O}$ is also known as a \emph{staircase} or, for $d=2$, as \emph{Ferrers diagram} or \emph{Young diagram}.
They are in one-to-one correspondence to (integer) partitions. 
\\
Similarly, hierarchical models in $d=3$ dimensions correspond to so-called plane partitions. Generating functions for partitions and plane partitions are known due to Euler and \cite{Ma12}, respectively.
No such functions are known for $d\ge 4$ \citep{OS99}. \\
Nevertheless, asymptotic results are known. Denote by $p_d(n)$ the number of order ideals with $n$ terms in $d$ dimensions.
Then, we have asymptotically $p_d(n)=\Theta(\exp (n^\frac{d-1}{d}))$, i.e. the number of hierarchical models grows
sub-exponentially \citep{BPA97}. Recall that the Big Theta notation $f(n)=\Theta(g(n))$ means that the function $f$ is bound from above and below by $g$ asymptotically. 
However, for $n$ fixed, $p_d(n)$ is polynomial in $d$ of degree $n-1$ \citep{ABMM67}.

\begin{definition}
	A term $t=X^{\alpha}$ evaluated at $D$ gives a \emph{design vector} $t(D)=(t(p_1), \ldots , t(p_n))^t$.
	These design vectors form the columns of the \emph{design matrix} $X_{\mathcal{O}}(D):=(X^{\alpha}(p))_{p\in D, \alpha \in \log \mathcal{O}}$.
\end{definition}

Note that the design matrix depends on the model $\mathcal{O}$ and the design $D$, and it is only defined up to a permutation of the terms in $\mathcal{O}$. 

\begin{example} Consider the 2-dimensional design $D=\{(1,-1), (-1,1), (-1, \allowbreak -1), (0,0)\}$. The design vector of e.g. the term $t=X^{(1,1)}=X_1X_2$ (interaction between $X_1$ and $X_2$) is $t(D)=(-1,-1,1,0)^t$. 
	The design matrix for the hierarchical model $\mathcal{O}=\{1,X_1,X_2,X_1X_2\}$ is
	\begin{equation*} \label{simple-example:dmatrix}
	X_{\mathcal{O}}(D) = \left(\begin{array}{rrrr}
	1 & X_1&X_2&X_1X_2 \\
	\hline
	1 &  1 &-1 &-1 \\
	1 & -1 & 1 &-1  \\
	1 & -1 &-1 & 1  \\
	1 &  0 & 0 & 0  
	\end{array}\right).\end{equation*}
	The respective regression model would be described by
	\begin{equation*}
	E(Y|X)=\beta_0 + \beta _1 X_1 + \beta_2 X_2 + \beta _3 X_1X_2
	\end{equation*}
	where $\beta=(\beta_0, \beta _1, \beta_2, \beta _3)^T$ in $\mathbb{R}^4$ is the unknown parameter vector to be estimated by the least squares method.
\end{example}	

In fitting regression models the problem of non-identifiability occurs if least-squares-estimates are not unique.
We next provide an algebraic definition for the set of identifiable  models with respect to a given design.
\begin{definition}
	A model is \emph{identifiable} if the \emph{design matrix} $X_{\mathcal{O}}(D)$ is of full rank. 
	The \emph{statistical fan} $S(D)$ of a design is the set of  hierarchical models identifiable by the design with as many terms as distinct design points.
\end{definition}
	In other words, the statistical fan is the set of all maximal identifiable order ideals.
\begin{example} Consider again the 2-dimensional design $D=\{(1,-1), (-1,1), (-1,-1), (0,0)\}$. Its statistical fan $S(D)$ contains three models, namely
$\mathcal{O}_3=\{1,X_1,X_2,X_1X_2\}$, $\mathcal{O}_2=\{1,X_1,X_2,X_1^2\}$, and $\mathcal{O}_3=\{1,X_1,X_2,X_2^2\}$. 
Figure \ref{fig:Ferrers} displays the Ferrers diagrams of the order ideals $\mathcal{O}_1,\mathcal{O}_2 , \mathcal{O}_3$.
The marked points all lie in the lattice $\mathbb{N}^2 \subset \mathbb{R}^2$ and they represent exponents, i.e. discrete logarithms, of bivariate terms, e.g. $(2,0)$ stands for $X^{(2,0)}=X_1^2$. The lattice points on the abscissa represent powers of $X_1$
	and the lattice points on the ordinate powers of $X_2$.
\end{example}
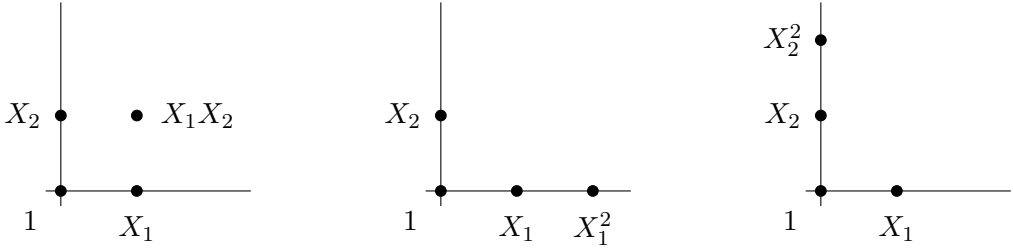
\begin{figure}[h!]
	\centering
	\begin{tikzpicture}[scale= 1]
	
	\draw [shift={(0 cm,0 cm)}] (-0.2,0) -- (2.5,0);
	\draw [shift={(0 cm,0 cm)}] (0,-0.2) -- (0,2.5);
	\filldraw [shift={(0 cm,0 cm)}] (0,0) circle (2pt)(0,1) circle (2pt)(1,0) circle (2pt)(1,1) circle (2pt);
	\draw [shift={(0 cm,0 cm)}] (-0.4,-0.4) node {$1$}; 
	\draw [shift={(0 cm,0 cm)}] (-0.5,1) node {$X_2$}; 
	\draw [shift={(0 cm,0 cm)}] (1.8,1) node {$X_1X_2$};
	\draw [shift={(0 cm,0 cm)}] (1,-0.5) node {$X_1$};  
	
	\draw [shift={(5 cm,0 cm)}] (-0.2,0) -- (2.5,0);
	\draw [shift={(5 cm,0 cm)}] (0,-0.2) -- (0,2.5);
	\filldraw [shift={(5 cm,0 cm)}] (0,0) circle (2pt)(0,1) circle (2pt)(1,0) circle (2pt)(2,0) circle (2pt);
	\draw [shift={(5 cm,0 cm)}] (-0.4,-0.4) node {$1$}; 
	\draw [shift={(5 cm,0 cm)}] (-0.5,1) node {$X_2$}; 
	\draw [shift={(5 cm,0 cm)}]  (2,-0.5) node {$X_1^2$};
	\draw [shift={(5 cm,0 cm)}] (1,-0.5) node {$X_1$}; 
	
	\draw [shift={(10 cm,0 cm)}] (-0.2,0) -- (2.5,0);
	\draw [shift={(10 cm,0 cm)}] (0,-0.2) -- (0,2.5);
	\filldraw [shift={(10 cm,0 cm)}] (0,0) circle (2pt)(0,1) circle (2pt)(1,0) circle (2pt)(0,2) circle (2pt);
	\draw [shift={(10 cm,0 cm)}] (-0.4,-0.4) node {$1$}; 
	\draw [shift={(10 cm,0 cm)}] (-0.5,1) node {$X_2$}; 
	\draw [shift={(10 cm,0 cm)}] (-0.5,2) node {$X_2^2$};
	\draw [shift={(10 cm,0 cm)}] (1,-0.5) node {$X_1$}; 
	\end{tikzpicture}
	\caption{Ferrers diagrams of the order ideals $\mathcal{O}_1, \mathcal{O}_2 , \mathcal{O}_3$ with the associated monomials.}
	\label{fig:Ferrers}
\end{figure}

For a generic design $D$ with $|D|=n$, all models (with $n$ terms) are identifiable \citep{PRW00}. Hence, the size of the statistical fan is bounded sub-exponentially in $n$.

\subsection{Gr{\"o}bner bases and algebraic fan} \label{Algfan}

Let $\prec$ be a \emph{term order} on $\mathcal{T}$, i.e., a total ordering on $\mathcal{T}$ which is multiplicative and a well-ordering. 
For a non-zero polynomial $f \in R$, we denote by $LT(f)$ its \emph{leading term}, that is the greatest term occuring in $f$ with respect to $\prec$. For an ideal $I$ in $R$, let $LT(I)$ be the ideal generated by all $LT(f)$ with $f\in I$, formally $LT(I):=\langle LT(f) \mid f\in I \rangle $.
\begin{definition}
	A finite generating subset $G\subset I$ is called a \emph{Gr{\"o}bner basis} for $I$ w.r.t. $\prec$ if $LT(I)=\langle LT(g) \mid g\in G \rangle $.
	 A Gr{\"o}bner basis $G$ is \emph{reduced} if the following conditions hold
	 \begin{enumerate}
	 	\item  The leading coefficient of each polynomial $g \in G$ is 1.
	 	\item   For any two distinct $g_1, g_2 \in G$, no monomial occurring in $g_1$ is a multiple of $LT(g_2)$. 
	 \end{enumerate}
We use the notation $G_{\prec}(I)$ for the unique reduced Gr{\"o}bner basis of $I$ w.r.t. $\prec$.
\end{definition}

The set of all terms that are not divided by the leading terms of the Gr{\"o}bner basis $G=G_{\prec}(I)$ form an hierarchical
model  $\mathcal{O}$. The residue classes of these terms in $\mathcal{O}$ form a vector space basis of the quotient
ring $R/I$. We call $(\mathcal{O},G)$ a \emph{Gr{\"o}bner pair}  for the ideal $I$. \\
\begin{example} Consider the $2^2$-factorial design $D=\{(1,-1), (-1,1), (-1, \allowbreak -1), (1,1)\}$.
	Since the values of the $X_1$- and $X_2$-coordinate are restricted to $\pm 1$, we know that the vanishing ideal $I=I(D)$ is generated by
	$X_1^2-1=(X_1-1)(X_1+1)$ and $X_2^2-1$.
	Indeed, these two polynomials form the Gr{\"o}bner basis $G$ of $I$ for any term ordering $\prec$. The terms not divided by the leading terms
	$X_1^2$ and $X_2^2$ form the hierarchical model $\mathcal{O}=\{1, X_1, X_2, X_1X_2\}$.  
\end{example}
We next provide a definition of the set of all maximal identifiable hierarchical models that can be obtained using 
Gr{\"o}bner basis techniques.
\begin{definition}
	The \emph{algebraic fan} $A(D)$ of a design $D$ is the set of all hierarchical models $\mathcal{O}$ such that
	$(\mathcal{O}, G_{\prec}(I))$ is a Gr{\"o}bner pair for the design ideal $I=I(D)$ if we run through all
	term orders $\prec$.
\end{definition}
The algebraic fan $A(D)$ is a subset of the statistical fan $S(D)$ and is in general much smaller than the latter.
$A(D)$ cannot contain more elements than there are reduced Gr{\"o}bner bases for $I$. 
The number of distinct reduced Gr{\"o}bner bases is connected to so-called corner cuts \citep{OS99}
and is asymptotically of order $O(n^{2d(d-1)/d+1})$. 
This upper bound has been improved by \cite{Wa02} who showed that the order of magnitude of the number of corner cuts lies
between $n^{d-1}\log n$ and $(n\log n)^{d-1}$.
Hence, the size of $A(D)$ grows polynomially in $n=|D|$ for fixed dimension $d$. 
A polynomial time algorithm for the computation of $A(D)$ has been given in \citep{BOT03}. There the authors compute the so-called \emph{universal Gr{\"o}bner basis} which encompasses all reduced Gr{\"o}bner bases of $I(D)$.

\section{Numerical statistical fan}
\label{NumFan}
Measurements come with errors - so do ``design'' points that are results from measurements themselves.
We are interested in identifiable maximal models that do not depend on small perturbations of the design points.
Such ``stable'' models will also exhibit numerical stability. \\
A measure of instability of a system of linear equations $Ax=b$ is the \emph{condition number}
$c(A):=||A||\cdot ||A^{inv}||$ where $A^{inv}$ is the Moore-Penrose pseudoinverse of $A$ and $||\cdot ||$ indicates some matrix norm. We will use only the induced 2-norm as matrix norm. The condition number of a square matrix
is always at least 1. If it is much larger than 1 then the matrix is ill-conditioned.

\begin{example} \citep{RKR16} 
Consider the design $D=\{(1,1), \allowbreak  (1,-1.001), (-1,1), (-1,-1)\}$. 
Its algebraic and statistical fan
are identical, and $S(D)$ has two
identifiable models, namely $\mathcal{O}_1=\{1, X_1, X_2, X_1X_2\}$ and  $\mathcal{O}_2=\{1, X_1, X_2, X_2^2\}$
with corresponding design matrices 
\begin{equation*} \label{simple-example:matrix}
\resizebox{.95\hsize}{!}{$
X_{\mathcal{O}_1}(D) = 
\left(\begin{array}{rrrr}
1 & X_1&X_2&X_1X_2 \\
\hline
1 &  1 & 1 & 1 \\
1 &  1 &-1.001 &-1.001  \\
1 & -1 & 1 &-1  \\
1 & -1 &-1 & 1  \\
\end{array}\right), \,
X_{\mathcal{O}_2}(D) = 
\left(\begin{array}{rrrr}
1 & X_1&X_2&X_2^2 \\
\hline
1 &  1 & 1 & 1 \\
1 &  1 &-1.001 &1.002001  \\
1 & -1 & 1 &1  \\
1 & -1 &-1 & 1  \\
\end{array}\right).$}
\end{equation*}

For $X_{\mathcal{O}_1}(D)$ we get almost $c(X_{\mathcal{O}_1}(D))=1.0007$ and for $X_{\mathcal{O}_2}(D)$ we have $c(X_{\mathcal{O}_2}(D))=4001$.
Indeed, the design vectors $t_1(D)=(1,1,1,1)^t$ and $t_2(D)=(1,1.002001,1,1)^t$ (for $t_1=1$ and $t_2=X_2^2$) are very close to each other.
That can be explained by the fact that there is a close perturbed design $\tilde{D}$, namely the $2^2$ full
factorial design $\{(\pm 1, \pm 1)\}$ for which $t_1(\tilde{D})$ and $t_2(\tilde{D})$ coincide. 
Indeed, $S(\tilde{D})$ has only one leaf $\mathcal{O}_1=\{1, X_1, X_2, X_1X_2\}$.  
\end{example}
In general, we need to define \emph{numerical linear dependence} for design vectors with error.
Several notions of  numerical linear dependence have been suggested. Since linear dependence of design vectors is connected
to polynomials vanishing at the design, one approach is to bound the evaluation of normed polynomials by a real threshold parameter $\epsilon >0$.
This is done e.g. in \cite{HKPP09} and \cite{Li14}. We follow another approach that requires the knowledge of the tolerance on the
data uncertainty, i.e. a tolerance vector  \citep{St04, AFT08, Fa10, To09}. 

\subsection{Numerical dependence of empirical points} \label{numDep}
We need a proper definition of perturbations in order to capture the notion of a noisy design with errors.
\begin{definition}
	Let $D$ be a given design and $\delta=(\delta_1, \ldots, \delta_d)$ with $\delta_i \ge 0$ a given vector of componentwise tolerances.   The pair $(D,\delta )$ is called an \emph{empirical design}. A point $\tilde{p}\in \mathbb{R}^d$
	is an \emph{$\delta$-perturbation} of a point $p=(c_1, \ldots , c_d)$ in $\mathbb{R}^d$ if $|\tilde{c}_i - c_i| < \delta_i$ for all $i=1,\ldots,d$. 
	A design $\tilde{D}$ is a $\delta$-perturbation of the design $D$ if a one-to-one mapping between $D$ and $\tilde{D}$ exists, such that each point $\tilde{p}\in \tilde{D}$ is a $\delta$-perturbation of the corresponding point $p\in D$.
\end{definition}
We will also call an empirical design a noisy design or a set of points of limited precision.\\
Next we define numerical linear dependence of empirical design vectors through the existence of a perturbed design for which we have exact linear dependence.
\begin{definition} \label{def:numDep}
	Given a set $\mathcal{O}=\{t_1, \ldots, t_k\}$, an empirical design $(D,\delta)$, and a monomial $t$, the design vector $t(D)$ \emph{numerically depends}
	on $\mathcal{O}(D)=\{t_1(D), \ldots, t_k(D)\}$ if there exists a $\delta$-pertubation $\tilde{D}$ of $D$ such that $t(\tilde{D})$ depends linearly on $\mathcal{O}(\tilde{D})$, i.e.,
	the 
	residual $\rho(\tilde{D})$ of the least squares problem $X_{\mathcal{O}}(\tilde{D}) \tilde{a} \approx t(\tilde{D})$ is a zero vector. 
\end{definition}
We only know the evaluations at the design $D$, and we can only solve the least squares problem $X_{\mathcal{O}}(D)a \approx t(D)$ for $a$. Nevertheless, Proposition \ref{Fass10} allows us to infer results for a perturbed design $\tilde{D}$.
\begin{proposition} \label{Fass10} \citep{Fa10} If there exists a $\delta$-perturbation $\tilde{D}$ of $D$ such that $\rho(\tilde{D})=0$ for the residual of
	the least squares problem $X_{\mathcal{O}}(\tilde{D}) \tilde{a} \approx t(\tilde{D})$, then the residual vector $\rho(D)$ (of the least squares problem $X_{\mathcal{O}}(D)a \approx t(D)$) satisfies
	\begin{equation} \label{FassinoIneq} |\rho (D)| \le |I-X_{\mathcal{O}}(D) X^{inv}_{\mathcal{O}}(D)|\sum_{k=1}^{d} \delta _k |\partial_k t(D) 
	- X_{\partial_k\mathcal{O}}(D)a| +O(\delta_M^2)  
	\end{equation}	
	Here $\delta _M=\max _{i=1}^d \delta _i$ and $X^{inv}_{\mathcal{O}}(D)$ is the Moore-Penrose inverse of the design matrix, i.e. $X^{inv}_{\mathcal{O}}(D)=(X^t_{\mathcal{O}}(D)X_{\mathcal{O}}(D))^{-1}X^t_{\mathcal{O}}(D)$. 
	And $\partial_k t$ denotes the partial derivative of the term $t$ with respect to $X_k$. Hence, $\partial_k t$ is in general not a term since it is not necessarily normed (i.e. the coefficient may be $\ne 1$). Similarly, $\partial_k\mathcal{O}$ denotes the multiset of all $\partial t/ \partial X_k$ for $t$ in $\mathcal{O}$. Hence, $X_{\partial_k\mathcal{O}}(D)$ is the matrix
		$(\partial_kt(p))_{p \in D, t \in \mathcal{O}}$.
\end{proposition}
Inequality (\ref{FassinoIneq}) provides a sufficient condition for showing that $t(D)$ is numerically independent
of the columns of $X_{\mathcal{O}}(D)$.
If we drop the $O(\delta_M^2)$ term, the Fassino condition becomes a heuristical condition for numerical independence.

More precisely, given a set $\mathcal{O}=\{t_1, \ldots, t_k\}$, an empirical design $(D,\delta)$, and a monomial $t$, the design vector $t(D)$ is declared to be numerically independent of
$\{t_1(D), \ldots, t_k(D)\}$ if the residual $\rho(D)$ (of the least squares problem $X_{\mathcal{O}}(D)a \approx t(D)$) satisfies
\begin{equation} \label{heurFass}|\rho (D)|_i > \sum_{j=1}^{n} |I-X_{\mathcal{O}}(D) X^{inv}_{\mathcal{O}}(D)|_{ij}\sum_{k=1}^{d} \delta _k |\partial_k t(D) 
- \sum_{l=1}^{|\mathcal{O}|}X_{\partial_k\mathcal{O}}(D)_{jl}a_l|
\end{equation}
for at least one design point $p_i \in D$.
To prove numerical dependence it would be preferable to find a $\delta$-perturbation $\tilde{D}$ of $D$ such that $\rho(\tilde{D})=0$. This is done in the root finding algorithm of \cite{FT13}.
In the applications in Section \ref{SprayData} 
we only deploy the heuristical Fassino condition (\ref{heurFass}) which we use as a proxy for numerical dependence.
Recall that, even if the $O(\delta_M^2)$ term is negligible, then condition (\ref{heurFass}) is only a sufficient condition for 
numerical independence. Hence, by checking condition (\ref{heurFass}), we might declare numerical dependence when there is none, i.e. the residuals are all small, but there exists no $\delta$-perturbed design on which they vanish exactly. 
We regard this as a good property of the heuristical Fassino condition (\ref{heurFass}), because such numerically independent terms which are close to dependence (in the sense that the residuals are small) may lead to bad (high) condition numbers.
By deploying condition (\ref{heurFass}) we guarantee that the norm of the residual vector is above some lower bound. This leads to a flexible upper bound for the evaluation of polynomials (divided by some polynomial norm) which describe linear dependencies between design vectors. In this sense our approach incorporates the idea of a real threshold parameter $\epsilon >0$  for normed almost vanishing polynomials as in \cite{HKPP09,Li14}, but we do not have a fixed $\epsilon >0$ which has to be fine-tuned.
In our case it depends on the model $\mathcal{O}$ and the empirical design $(D, \delta)$. In particular, we use knowledge about the tolerance on the data uncertainty given by the vector $\delta$.

\subsection{Stable order ideals and numerical fans}\label{stOI}

The following definition captures an analogue of the notion of identifiability (of a model) in the context of noisy designs. 
\begin{definition} \label{def:stOI}
	Let $(D,\delta)$ be an empirical design. An order ideal $\mathcal{O}$ is called \emph{numerically stable} (or just \emph{stable}) if  the evaluation matrix $X_{\mathcal{O}}(\tilde{D})$ has full column rank for each $\delta$-perturbation $\tilde{D}$ of $D$.
\end{definition}

Note that the design (or evaluation) vectors of all monomials in a numerically stable order ideal $\mathcal{O}$ are numerically independent. \\
Let us  distinguish two concepts of maximality of stable order ideals, namely weakly maximal and maximal.
\begin{definition}
	Let $(D, \delta)$ be an empirical design. A stable order ideal $\mathcal{O}$ is called \emph{weakly maximal} if $t(D)$ is numerically dependent of $X_{\mathcal{O}}(D)$ for all $t$ in the corner of $\mathcal{O}$.
	A stable order ideal $\mathcal{O}$ is called \emph{maximal} if there exists no stable order ideal $\mathcal{O}'$ such that $\mathcal{O} \subset \mathcal{O}'$.
\end{definition}
It easy to show that both notions of maximality are equivalent.
\begin{proposition} \label{maximality}
	Let $(D, \delta)$ be an empirical design. A stable order ideal for $(D, \delta)$ is weakly maximal if and only if it is maximal. 
\end{proposition} 
{\bf Proof.} By definition, $\mathcal{O}$ being maximal implies $\mathcal{O}$ being weakly maximal.
Now, assume that $\mathcal{O}$ is weakly maximal. Every stable order ideal $\mathcal{O}'$ that is a proper superset of $\mathcal{O}$ necessarily contains an element in the corner set of $\mathcal{O}$ - otherwise $\mathcal{O}'$ wouldn't be an order ideal. But this is in contradiction to $\mathcal{O}$ being weakly maximal. Hence, there is no stable order ideal that properly contains $\mathcal{O}$, i.e. $\mathcal{O}$ is maximal.\quad $\Box$\\

Nevertheless, it will prove ueseful to distinguish both concepts of maximality when we move to heuristical notions of numerical dependence. The notion of maximal stable order ideals allows us to introduce a numerical analogue of the statistical fan of a design. 
\begin{definition}
	The set of all maximal stable order ideals of an empirical design $(D, \delta)$ is called the \emph{numerical statistical fan} $S_{num}(D,\delta)$ of the empirical design.
\end{definition}
In the next subsection we will tackle the problem of computing the numerical statistical fan of an empirical design.
First we describe a strategy to find some/any maximal stable order ideal \citep{To09} summarized in Algorithm \ref{alg:stOI} which is based on the following remark.

\begin{remark} \label{def:heurstOI}
	Let $(D, \delta)$ be a non-empty finite empirical design. $\{1\}$ is a stable order ideal (for $(D, \delta)$).
	Assume $\mathcal{O}$ is stable and $t$ is in the corner set of $\mathcal{O}$. Then $\mathcal{O}\cup \{t\}$ is stable if $t(D)$ is numerically independent of $\mathcal{O}(D)$.
\end{remark}

 Note that an order ideal $\mathcal{O}$ can be efficiently encoded by its maximal elements
(w.r.t. divisibilty) or equivalently by the minimal elements of its complement $\mathcal{T}\setminus \mathcal{O}$. These minimal elements
form the \emph{corner set} of $\mathcal{O}$. They are precisely the monomials $t$ such that $\mathcal{O} \cup \{t\}$ is again an order ideal. Therefore they are the new candidate elements to be included in $\mathcal{O}$ in a recursive computation of $\mathcal{O}$. This recursive computation of a maximal stable order ideal $\mathcal{O}$ is a modification of the Buchberger-M{\"o}ller algorithm \citep{MB82}. 

\begin{algorithm}
	\DontPrintSemicolon
	Initialize $\mathcal{O}=\{1\}$ and $C=\{\}$.\;
	 Choose a term $t$ in the corner set of $\mathcal{O}$ not belonging to $C$.
	If no such $t$ exists, return $\mathcal{O}$ and stop. \;
	If $t(D)$ is numerically linear independent of $\mathcal{O}(D)$ then add $t$ to $\mathcal{O}$. Otherwise, add $t$ to $C$. \;
	Go to step 2. \;
	\caption{MaximalStableOrderIdeal()}
	\label{alg:stOI}
\end{algorithm}
\begin{theorem} \label{thm:stOI}
	Let $(D,\delta)$ be an empirical design with $|D|$ finite. Algorithm \ref{alg:stOI} terminates after finitely many steps and returns a maximal stable order ideal $\mathcal{O}$ for the empirical design $(D, \delta)$.
\end{theorem}
{\bf Proof.} First we show that the algorithm terminates. Since the corner set of any finite subset of $\mathbb{N}$ is finite, the algorithm terminates if and only if $\mathcal{O}$ is finite. But the size of a stable order ideal $\mathcal{O}$ is bounded above by $|D|$. \\
 $\mathcal{O}$ is obviously stable by construction.
We prove that that the returned set of terms $\mathcal{O}$ is indeed an order ideal by induction over the cardinality of $\mathcal{O}$. The initialized order ideal $\mathcal{O}=\{1\}$ forms the basis of induction.
For the induction step, assume that $\mathcal{O}$ is an order ideal. The next term $t \in \mathcal{T} \setminus\mathcal{O}$ added to $\mathcal{O}$
is from the corner set of $\mathcal{O}$. Assume that $\mathcal{O} \cup \{t\}$ is not an order ideal, then there exists a $t' \in  \mathcal{T} \setminus\mathcal{O}$ such that $t'\mid t$ and $t'\ne t$. This in contradiction to $t$ being a minimal element (w.r.t. divisibility) from $\mathcal{T} \setminus\mathcal{O}$. \\
Algorithm \ref{alg:stOI} stops when $C$ coincides with the corner set of $\mathcal{O}$. Since the design vectors of the elements in $C$ are numerically dependent of $\mathcal{O}(D)$, this means that $\mathcal{O}$ is weakly maximal. By Proposition \ref{maximality} this implies maximality.
\quad $\Box$\\
Note that any strategy that chooses a candidate term $t$ in the corner set of $\mathcal{O}$ can be applied. 
In the Buchberger-M{\"o}ller algorithm \citep{MB82} - as well as its numerical analogue, the numerical Buchberger-M{\"o}ller (NBM) algorithm \citep{Fa10} - the monomial $t$ is chosen as the smallest candidate w.r.t. a fixed monomial ordering $\prec$.
Indeed, given an empirical design $(D, \delta)$ and a term ordering $\prec$, the NBM algorithm returns a stable order ideal
$\mathcal{O}$ and a set $G$ of of almost vanishing polynomials on $D$. The output $(\mathcal{O},G)$ from the NBM algorithm is the numerical analogue to the Gr{\"o}bner pair provided by the classical M{\"o}ller-Buchberger algorithm.
Given a tolerance vector $\delta $, numerical linear dependence is checked in the NBM algorithm \citep{Fa10} by the heuristic Fassino condition.

\begin{definition} \citep{RKR16} Let $(D, \delta)$ be an empirical design.
	The \emph{numerical algebraic fan} $A_{num}(D,\delta)$ is the set of all stable order ideals $\mathcal{O}=\mathcal{O}_{\prec}$ such that 
	$(\mathcal{O}_{\prec}, G_{\prec})$ is the output of an NBM algorithm, running through all term orderings $\prec$. 
\end{definition}
By defintion, the numerical algebraic fan $A_{num}(D,\delta)$ is a subset of the numerical statistical fan $S_{num}(D,\delta)$.
For $\delta =0$, we have $A_{num}(D,\delta)= A(D)$ and $S_{num}(D,\delta)=S(D)$. 
We are interested in the numerical statistical fan $S_{num}(D,\delta)$ since it gives us all maximal stable order ideals. 
The numerical algebraic fan $A_{num}(D,\delta)$ may be an important subset of $S_{num}(D,\delta)$ to consider, if the numerical statistical fan is not feasible to compute. This is not the case in our applications. 
Note that the computation of the numerical algebraic fan has not been implemented yet. In \cite{RKR16} 
only a subset of $A_{num}(D,\delta)$ is computed by chosing three popular term orderings and permuting the coordinates.

\subsection{Computation of the statistical fan} \label{CompSF}
Hashemi, Kreuzer and Pourkhajouei (2019) suggest a recursive algorithm to compute all border bases of a finite 0-dimensional ideal $I=I(D)$. This algorithm necessarily also computes all maximal order ideals and can thus be utilized to get the statistical fan $S(D)$. 
The following algorithms are modifications of Algorithms 3 and 4 from \cite{HKP19}.
The function AllOrderIdeals() (see Algorithm \ref{alg:AllOIs}) computes  a set $S$ of all maximal order ideals for a design $D$. It simply initializes an empty set $S$, an empty order ideal $\mathcal{O}$ and its corresponding design matrix $M$, and calls the main function AllOIStep. 

\begin{algorithm}
	Initialize $S:=\{\}$; $\mathcal{O}=\{\}$ (empty sets)\;
	Initialize $M$ as $n\times 0$-matrix\;
	$S, \mathcal{O}, M:=$ AllOIstep$(S, \mathcal{O}, M)$\;
	\caption{AllOrderIdeals()}
	\label{alg:AllOIs}
\end{algorithm}

The main function AllOIStep($S$, $\mathcal{O}$ , $M$) (see Algorithm \ref{alg:AllOIstep}) changes the set $S$, the current order ideal
$\mathcal{O}$ and its corresponding design matrix $M=X_{\mathcal{O}}(D)$.
It adds to the set $S$ all maximal order ideals that properly contain the current order ideal $\mathcal{O}$.
An order ideal $\mathcal{O}$ is added to the set $S$ if it has maximal size $n=|D|$ and was not added earlier.
If $\mathcal{O}$ is not maximal, we check for all terms $t$ in the corner set $C$ of $\mathcal{O}$
whether the design vector $t(D)$ is linear independent of the columns of the design matrix $M$.
In the case of linear independence the function calls itself to add all maximal order ideals that properly contain  $\mathcal{O} \cup \{t\}$.
Since $\mathcal{O}$ and $M$ are updated and we run through all $t \in C$, we have to store the original $\mathcal{O}_{in}:=\mathcal{O}$ (and $M_{in}:=M$) that has
corner set $C$. 

\begin{algorithm}
	$\mathcal{O}_{in}:=\mathcal{O}$; $M_{in}:=M$\;
	\lIf{$|\mathcal{O}|=n$ {\bf and} $\mathcal{O} \notin S$}{include $\mathcal{O}$ in  $S$}
	\If{$|\mathcal{O}|< n$}{
		$C:=$ set of all terms $t \notin\mathcal{O}$ s.t. $\mathcal{O} \cup \{t\}$ is an order ideal\;
		\For{$t$ {\bf in} $C$}{
			$\mathcal{O}:=\mathcal{O}_{in}$; $M:=M_{in}$\;
			\If{$t(D)$ {\rm is linear independent of columns of} $M$}{ 
				Add $t(D)$ as last column to $M$\;
				$S$, $\mathcal{O}$, M:=AllOIStep($S$,$\mathcal{O} \cup \{t\} , M$)\;
			}
		}
	}
	{\bf return} $S$, $\mathcal{O} , M$\; 
	\caption{AllOIStep($S$, $\mathcal{O} , M$)}\label{alg:AllOIstep}
\end{algorithm}

\begin{proposition} \label{thm:AllOIstep}
	Initialize the empty set $S=\{\}$. Let $\mathcal{O}$ be an order ideal for $D$.
	Algorithm AllOIStep($S$, $\mathcal{O} , X_{\mathcal{O}}(D)$ terminates after finitely many steps and computes the set $S$ of all maximal order ideals for $D$ which properly include $\mathcal{O}$.
\end{proposition}

{\bf Proof.} This is a special case of Proposition \ref{thm:AllstOIstep}. \quad $\Box$
\begin{theorem}
	Let $D$ be a finite design. Algorithm \ref{alg:AllOIs} computes the  statistical fan $S(D)$.
\end{theorem}

{\bf Proof.} According to Proposition \ref{thm:AllOIstep} Algorithm \ref{alg:AllOIs} computes the set $S$ of all maximal order ideals for $D$ that properly contain the empty order ideal $\{\}$, i.e. the full statistical fan $S(D)$.
\quad $\Box$\\

Note that in \cite{HKP19} the design vector $t(D)$ was first reduced with respect to the columns of $M$ before it was added as a new column.
Reduction is not needed for the computation of the statistical fan. Furthermore, as we will see in the next section, it is not wanted for finding the numerical statistical fan as the design vectors $t(D)$ (the columns of $M$) are itself of interest.

\subsection{Computation of the numerical statistical fan} \label{CompNF}
In order to compute the numerical fan $S_{num}(D,\delta)$ we adapt Algorithm \ref{alg:AllOIstep}
by replacing linear independence by numerical linear independence. However, this is not sufficient.
It is clear that Algorithms \ref{alg:AllOIs} and \ref{alg:AllOIstep}  cannot
be used directly to compute the numerical statistical fan since the breaking condition in Algorithm \ref{alg:AllOIstep} is $|\mathcal{O}|=n$ ($n=|D|$), but maximal stable order ideals might have  $|\mathcal{O}|<n$. 

\begin{algorithm}
	Initialize $S:=\{\}$; $HS:=\{\}$; $\mathcal{O}:=\{\}$   (empty sets) \;
	Initialize $C:=\{1\}$ \;
	Initialize $M$ as $n\times 0$-matrix \;
	$S, \mathcal{O}, C, M, HS:$= AllStableOIstep($S, \mathcal{O}, C, M, HS)$ \;
	{\bf return} $S$	
	\caption{AllStableOrderIdeals()}
	\label{alg:AllstOIs}
\end{algorithm}

Algorithm \ref{alg:AllstOIstep}
is a function that calls itself in order to compute all maximal stable order ideals that properly contain a given order ideal $\mathcal{O}$. \\
Improving Algorithm \ref{alg:AllOIstep}, here we also update the corner set $C$ of $\mathcal{O}$ and a set $HS$ of all stable order ideals (that properly contain a given order ideal $\mathcal{O}$).
This has two benefits. First, it is much faster to compute the corner set of $\mathcal{O} \cup \{t\}$ from $C$ - the corner set of $\mathcal{O}$ -
rather than from $\mathcal{O} \cup \{t\}$ directly. Second, by using a set of all order ideals, we avoid visiting an order ideal several times. 
Here we can replace the set $HS$ of order ideals by a set of corner sets for efficiency. 
Furthermore, it can be replaced by a hash set (therefore $HS$) of hash values, because we are not interested in all order ideals - only in the (weakly) maximal stable ones. 

\begin{algorithm}
	$\mathcal{O}_{in}:=\mathcal{O}$; $C_{in}:=C$; $M_{in}:=M$ \;
	\If{ $\mathcal{O} \notin S$}{
		$Maxbool:=TRUE$\;
		\For{$t$ {\bf in} $C_{in}$}{
			$\mathcal{O}:=\mathcal{O}_{in}$; $C:=C_{in}$; $M:=M_{in}$ \;
			\If{$t(D)$ {\rm is num. linear independent of the columns of} $M$}{ 
				$Maxbool:=FALSE$\;
				\If{$\mathcal{O} \notin HS$}{
					Add $t(D)$ as last column to $M$ \;
					$\mathcal{O}:=\mathcal{O} \cup \{t\}$  \;
					$C$:= corner set of $\mathcal{O} \cup \{t\}$\; 
					Include $\mathcal{O}$ in the set $HS$ \;            
					$S$, $\mathcal{O}, C, M$, $HS$:=AllOIStep($S$,$\mathcal{O},C, M$, $HS$) \;
				}
			}
		}
		\lIf{Maxbool == TRUE}{Include $\mathcal{O}$ to the set $S$}
	}
	{\bf return} $S$, $\mathcal{O} , C, M, HS$ \; 
	\caption{AllStableOIStep($S$, $\mathcal{O}, C, M$, $HS$)}\label{alg:AllstOIstep}
\end{algorithm}

An stable order ideal is added to the set $S$ if $Maxbool == TRUE$, i.e.  if for all $t \in C$, $t(D)$ is numerically linear dependent of
the columns of $M$. That is the definition of a weakly maximal stable order ideal.

\begin{proposition} \label{thm:AllstOIstep}
	Initialize empty sets $S=\{\}$, $HS=\{\}$. Let $\mathcal{O}_0$ be a fixed stable order ideal for the finite empirical design $(D, \delta )$ and $C_0$ its corner set.
	 On input ($S$, $\mathcal{O}_0, C_0,X_{\mathcal{O}_0}(D)$, $HS$) Algorithm \ref{alg:AllstOIstep} terminates after finitely many steps and computes the set $S$ of all maximal stable order ideals for $(D, \delta )$ which properly contain $\mathcal{O}_0$. Furthermore, it also computes the set $HS$ of all stable order ideals for $(D, \delta )$ which properly contain $\mathcal{O}_0$. 
\end{proposition}
{\bf Proof.}
By construction, every set $\mathcal{O}$ that appears during the computation will be a stable order ideal.
In particular, the proof for the order ideal property is as in the proof of Theorem \ref{thm:stOI}. \\
 Let $A$ be the set of all stable order ideals for $(D, \delta )$ which properly contain $\mathcal{O}_0$.
 $A$ can be decomposed as disjoint union $\biguplus_{i} A_i$, $i=|\mathcal{O}_0|+1, \ldots , |D|$, with $A_i$ containing order ideals in $A$ of cardinality $i$ only.
 We show that Algorithm \ref{alg:AllstOIstep} finds any element in $A$ by induction over $i$.
 Assume by induction hypothesis that Algorithm \ref{thm:AllstOIstep} has added all elements from $A_i$ to $HS$.
 Let $\mathcal{O}'$ be an arbitrary element in $A_{i+1}$. Then $\mathcal{O}'=\mathcal{O} \uplus \{t\}$ for some $\mathcal{O} \in A_i$ and $t$ is simultaneously a maximal element (w.r.t. divisibility) of $\mathcal{O}'$ and in the corner set of $\mathcal{O}$. Since $\mathcal{O}$ was added to $HS$ by induction hypothesis, Algorithm \ref{alg:AllstOIstep} went through the corner set of $\mathcal{O}$ and added also $\mathcal{O}$' (if it was not already previously added) to $HS$. \\
 Since we compute all elements of $A$, we also compute all weakly maximal ones. Only weakly maximal (i.e. maximal) stable order ideals are added to the set $S$. \\ 
  It remains to show that the algorithm terminates. Observe that the function  AllStableOIStep($S$, $\mathcal{O}, C, M$, $HC$) calls itself whenever we add a new order ideal $\mathcal{O}$ to $HS$, i.e. the number of function calls is $|HS|=|A|$. For each order ideal $\mathcal{O}$ (added to $HS$) we run through its finite corner set.
  Since $A$ is finite, the number of steps in the whole algorithm is finite.
\quad $\Box$

\begin{theorem}
	Let $(D, \delta )$ be a finite empirical design. Algorithm \ref{alg:AllstOIs} computes  the numerical statistical fan $S_{num}(D, \delta)$.
\end{theorem}

{\bf Proof.} According to Theorem \ref{thm:AllstOIstep} Algorithm \ref{alg:AllstOIs} computes the set $S$ of all maximal stable order ideals for $(D, \delta )$ that contain the empty order ideal $\{\}$, i.e. the full numerical statistical fan $S_{num}(D, \delta)$.
\quad $\Box$\\

\subsubsection{Numerical fan and Fassino condition} \label{NumfanFass}

Numerical linear independence can be empirically checked by deploying the heuristical Fassino condition.
Since this condition is not sufficient for true numerical dependence (as defined in Definition \ref{def:numDep}), it might provide some false positives.
Even more troublesome, the notion of stable order ideal is then not well-defined anymore. 

Heuristically, stability of order ideals can be checked recursively as in Remark \ref{def:heurstOI} by using condition (2). 
It is argued at  the end of section 4 in \cite{Fa10} that there is a high probability that if $\mathcal{O}$ is stable w.r.t.  Remark \ref{def:heurstOI} (using the Fassino condition) then it is stable (according to Definition \ref{def:stOI}).  \\
However, the two concepts of maximality of stable order ideals (defined in section \ref{stOI}) are then not equivalent anymore. 
If $\mathcal{O}$ is maximal, then it is weakly maximal. But the opposite may not hold
when we check numerical dependence by the heuristical Fassino condition.
To deal with this problem we use the Fassino condition in Algorithm \ref{alg:AllstOIs}, but then eliminate all order ideals which are not maximal.
Hence, Algorithm \ref{alg:AllstOIs} first computes a set $S$ of all weakly maximal order ideals using as main function Algorithm \ref{alg:AllstOIstep}.
Then, in order to get the numerical statistical fan of maximal stable order ideals, one has to check for inclusions
$\mathcal{O} \subsetneq \mathcal{O}'$ in the set $S$, and we keep
only the maximal order ideals w.r.t. inclusion. \\
Despite the conceptual difficulties when using the heuristic Fassino condition to check for numerical dependence,
it provides the some nice invariance properties. 
\begin{remark} \label{FassInvar}
	Let $(D, \delta)$ be an empirical design in $K^d$.  We write $S_{num}^{Fas}(D, \delta)$ for the numerical statistical fan which was computed by deploying the heuristical Fassino condition. \\
	(1) Let $p \in K^d$ be a fixed translation vector defining a translated design $D+p=\{q+p \in K^d \mid q \in D \}$. Then the translated empirical design $(D+p, \delta)$ has the same numerical statistical fan, i.e.
    \[  S_{num}^{Fas}(D, \delta)=S_{num}^{Fas}(D+p, \delta ). \]
    (2) Let $s \in K^d$ be a fixed scaling vector defining a componentwise scaled empirical design $(D', \delta')$ by $D'=\{(s_1c_1, \ldots, s_dc_d)\mid (c_1, \ldots, c_d) \in D \}$ and $\delta'=(|s_1|\delta_1, \ldots, |s_d|\delta_d)$. Then the componentwise scaled empirical design has the same numerical statistical fan, i.e.   \[  S_{num}^{Fas}(D, \delta)=S_{num}^{Fas}(D', \delta '). \]
\end{remark}

Remark \ref{FassInvar} is a straightforward consequence of Theorem 4.1 in \cite{Fa10} which states the same invariance w.r.t. scaling and translation for the output order ideal $\mathcal{O}$ of the NBM algorithm. Obviously, the proof of Theorem 4.1 in \cite{Fa10} does not use the given term ordering, and the invariance is a property of the Fassino condition. 

\subsection{Examples}
In this subsection we apply the algorithms from Sections \ref{CompSF} and \ref{CompNF} to compute the statistical fan and the numerical statistical fan for some small examples of (empirical) designs.
\begin{example} Consider again the design $D=\{(1,1), (1,-1.001), (-1,1), \allowbreak (-1,-1)\}$ from the beginning of Section \ref{NumFan}. 
Its statistical fan $S(D)$ has two
identifiable models, namely $\mathcal{O}_1=\{1, X_1, X_2, X_1X_2\}$ and  $\mathcal{O}_2=\{1, X_1, 
X_2, X_2^2\}$. This design lies very close to the $2^2$-full factorial design $\tilde{D}=\{(1,1), (1,-1), \allowbreak (-1,1), (-1,-1)\}$ which famously has only $\mathcal{O}_1$ in its statistical fan. Hence the numerical statistical fan of the noisy design $(D, \delta)$ should also only contain $\mathcal{O}_1$ for any tolerance vector $\delta $ with sufficiently big $|| \delta ||$. Indeed, applying Algorithm \ref{alg:AllstOIs} to $(D, \delta )$, we get
$S_{num}(D, \delta )={\mathcal{O}_1}$ for all $\delta=(0, \delta_{X_2})$ with $\delta_{X_2} \ge 1/2000=0.0005$.
Note that here the critical value of $||\delta ||$ is roughly the range of each component divided by  $c(X_{\mathcal{O}_2}(D))=4001$. \\
\end{example}
The next example compares Algorithm \ref{alg:AllstOIs} with the NBM algorithm \citep{Fa10}.
\begin{example} (Example 6.4 in \cite{Fa10}).
Consider the empirical design  $(D, \delta )$ with $D=\{(1,6),(2,3),(2.449, 2.449),(3,2),(6,1)\}$ and
$\delta =0.018\cdot(1,1)$. The NBM algorithm (with term order DegLex) computes only the
stable order ideal $\mathcal{O}=\{1,X_1,X_2,X_2^2,X_2^3\}$. Our algorithm  \ref{alg:AllstOIs} computes the numerical statistical fan $S_{num}(D, \delta )=
\{\{X_1^2,X_2^2\},\{X_1^3, X_2\},\\ \{X_1,X_2^3\},
\{X_1^4\},\{X_2^4 \}\}$ where we displayed for each order ideal only its maximal elements, 
	e.g.  $\{X_1^2,X_2^2\}$ stands for the order ideal $\{1, X_1, X_2, X_1^2, \allowbreak X_1X_2, X_2^2\}$, and $\{X_1,X_2^3\}$ represents the order ideal $\mathcal{O}$ computed by the NBM algorithm. 
\end{example}
While the NBM algorithm computes only one stable order ideal (for a given term order), our Algorithm \ref{alg:AllstOIs} computes them all. Even if we would run the NBM algorithm through all possible term orderings - which is a difficult task to implement - we could only compute the numerical algebraic fan $A_{num}(D, \delta)$ \citep{RKR16} which is a subset of $S_{num}(D, \delta)$.

\section{Application to thermal spraying data}
\label{SprayData}
We are going to compute the numerical statistical fan for real data coming from a High Velocity Oxygen Fuel (HVOF) thermal spraying process. Section \ref{Data} describes the data and the designs $D_X$ and $D_Y$ in question. 
In Section \ref{Sfan} we compute the numerical statistical fan $S_{num}^{Fas}(D_Y,\delta)$ and compare it to the 
statistical fan $S(D_Y)$.\\
All computations were performed using the computer algebra system MAGMA \citep{BCP97}.

\subsection{Data}
\label{Data}
In an HVOF process cermet powder particles are sprayed by a spraying gun to build a coating on a specimen. Controllable machine parameters ($X$ variables) are
varied according to an experimental design $D_X$. In-flight properties ($Y$ variables) are measured during the process.
Coating properties ($Z$ variables) are very time-consuming and expensive to measure
as the specimen has to be destroyed. Thus it is desirable to predict coating properties on the basis of particle properties, i.e., we are considering models $Z=Z(Y)$.
Table \ref{tab:parameters} displays all $X$, $Y$, and $Z$ variables involved. 

\begin{table}[t!]
	\caption{Controllable and measured variables in the spraying process}
	\label{tab:parameters}
\vskip .2cm
\centerline{\tabcolsep=3truept
	\begin{tabular}{|l|l|l|}
		\hline
		process parameters $X$ &  in-flight properties $Y$ & coating properties $Z$ \\
		\hline
		Kerosene & Temperature ($T$) &  Porosity  \\
		Lambda  & Velocity ($V$) &  Hardness \\
		Stand-off Distance  & Flame width ($W$) & Thickness \\
		Feeder Disc Velocity  & Flame intensity ($I$) &  Deposition efficiency \\
		\hline
	\end{tabular}}
\end{table}

\begin{figure}[htbp]
	\centering
	\includegraphics[trim=0cm 0.5cm 0cm 0.5cm,width=1\textwidth]{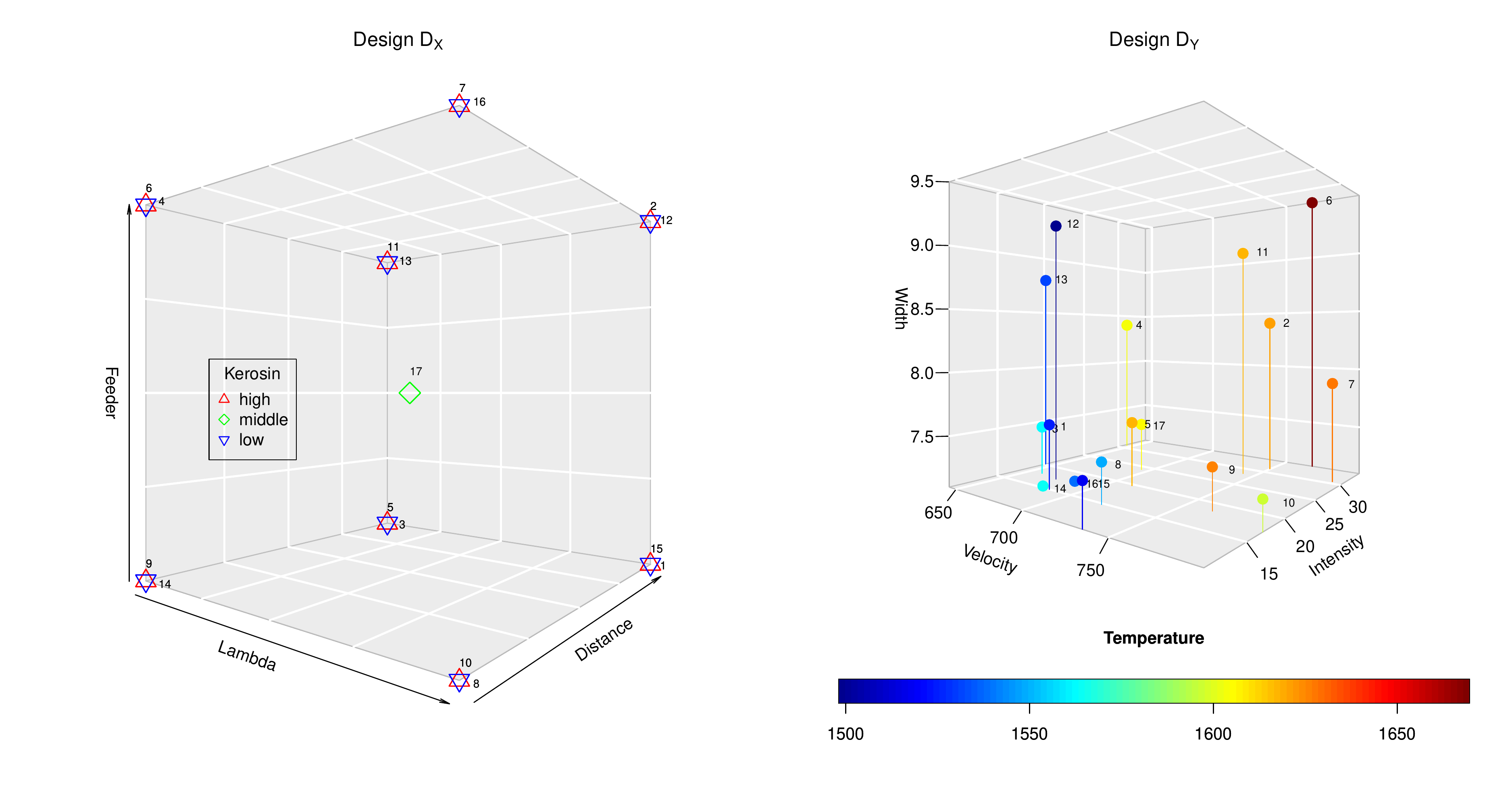}
	\caption{Designs $D_X$ and $D_Y$ with corresponding design points enumerated.}
	\label{fig:noisydesign}
\end{figure}

While the other variables are self-explanatory,
Lambda is defined as the quotient of actual oxygen-fuel mass ratio to its stoichiometric value.
The experimental design $D_x$ is chosen as a $2^4$ factorial design with one center point, i.e., $D_X=\{\pm 1\}^4 \cup \{0 \in\mathbb{R}^4\}$. The measured values of in-flight properties form a noisy design $D_Y \subset \mathbb{R}^4$ with
$|D_Y|=|D_X|=17$.
Figure \ref{fig:noisydesign} displays the design $D_X$ and the corresponding values 
in the design $D_Y$. Since we can only draw a perspective plot of 3 dimensions, the 4-th dimension in these 4-dimensional designs $D_X, D_Y$ was either displayed by different symbols for different levels (for kerosene in $D_X$) or by using a continous color spectrum (for temperture in $D_Y$). The numbers $i=1, \ldots , 17$ label corresponding design points $p_i \in D_X$ and $q_i\in D_Y$, i.e. the data point $q_i \in D_Y$ is the result from a measurement
with process parameters given by $p_i \in D_X$. 

\subsection{Computation of the statistical fans}
\label{Sfan}
Since the system which measures the particle and flame properties records data every second, we were able to estimate a bound for the standard deviation for every data point and in-flight property. We took the maximum over all data points (for each in-flight property) and chose 
$\delta _i=2\cdot \sigma _i^{(\max)}$ for $i\in (T,V,W,I)$.
This procedure led to the tolerance vector $\delta=(\delta _T, \delta _V, \delta _W, \delta _I)=(12.5,7,0.3, 1.5)$.
We check that all empirical points of the noisy design $(\mathcal{D}_Y, \delta)$ are well seperated for this choice of $\delta$ which is indeed the case.
If this were not the case we could  first apply a data reduction step where we merge points that are contained in each others $\delta$-boxes. Explicit algorithms for such data reduction are given in  \cite{To09}. \\  
We compute the numerical statistical fan $S_{num}^{Fas}(\mathcal{D}_Y, \delta)$ of the noisy design $(\mathcal{D}_Y, \delta)$
using Algorithms \ref{alg:AllstOIs} and \ref{alg:AllstOIstep} together with the additional eliminations as outlined in subsection \ref{NumfanFass}.
We find that $S_{num}^{Fas}(\mathcal{D}_Y, \delta)$ has 45 maximal order ideals of different cardinality, i.e. the hierarchical models vary in number of effects. \\
To explore the resulting models further, we look at the condition numbers of the design matrices as an alternative characteristic of stability. Since the condition number is not invariant under scaling and translation of the coordinates of the design points, we standardize our design $D_Y$ such that the range of each in-flight property is exactly $[-1,1]$. This makes no difference when computing the numerical fan since the heuristical Fassino condition is invariant under scaling and translation of each coordinate according to Remark \ref{FassInvar}.
The condition numbers of the design matrices turn out to be reasonably small. The largest condition number is 
$c(X_{\mathcal{O}}(D))=62.25$  for $\mathcal{O}=\{ T^2V, V^5 \}$ (only maximal elements displayed), and the smallest
condition number is $c(X_{\mathcal{O'}}(D))=5.65$ for  $\mathcal{O}=\{ TW, V^2, W^2 \}$. For models with high degree terms as $V^5$ we expect higher condition numbers. The reason why we cannot get close to 1 are design points which are well separated by $\delta$ but still quite close to each other (take a look at $D_Y$ in Figure \ref{fig:noisydesign}). \\ 
The next step in a statistical data analysis would be to use these hierarchical models in a model selection procedure. 
Here, we focus on some properties of our approach instead.
One question might be how the size of the numerical statistical fan changes when the $\delta$-vector becomes smaller.	
Table \ref{scaledEps}  displays the size  of the numerical statistical fan $S_{num}^{Fas}(D_Y, k\delta)$ for  $\delta=(12.5, 7, 0.3, 1.5)$
and different scale factors $k$. Additionally, we also give the number of all stable order ideals and all weakly maximal among them. 

\begin{table}[t!]
		\caption{Number of (maximal) stable order ideals of the empirical design $(D_Y, k\delta)$ for $\delta =(\delta _T, \delta _V, \delta _W, \delta _I)=(12.5, 7, 0.3, 1.5)$ and different scales $k$}
	\label{scaledEps}
\vskip .2cm
\centerline{\tabcolsep=3truept
	\begin{tabular}{|c|c|c|c|}
		\hline
		scale $k$ & $|S_{num}^{Fas}(D_Y, k\delta)|$ & \#\{\rm weakly max. OIs\} & \#{\rm stable order ideals}  \\
		\hline		
		2      &     5  & 10  &  30 \\
		1.5    &    11  & 22  & 97 \\
		1.2    &    25  & 44  & 210  \\					
		1      &    45  & 68  & 481 \\	
		0.9     &    77  & 103 & 777 \\
		0.8     &    165 & 230 & 1551 \\	   
		0.7     &    342 & 511 & 3079 \\	 
		0.6     &    697 & 974 & 6740 \\
		0.5     &    1488& 2086& 16233 \\
		\hline
	\end{tabular}}
\end{table}
As expected $|S_{num}^{Fas}(D_Y,\delta)|$ becomes larger for smaller scale factor $k$, because numerical aliasing becomes less likely for smaller
$\delta$-boxes. \\
To get an impression of which sizes of the numerical statistical fan occur, we display the number of occurences of maximal stable order ideals with different cardinalities in Table \ref{tab:occur}.

\begin{table}[t!]
		\caption{Maximal Stable order ideals by cardinalty in numerical statistical fan $S_{num}^{Fas}(D_Y, k\delta)$ for $\delta =(\delta _T, \delta _V, \delta _W, \delta _I)=(12.5, 7, 0.3, 1.5)$ and different scales $k$.}
	\label{tab:occur}
	\vskip .2cm
	\centerline{\tabcolsep=3truept
	\begin{tabular}{|c|cc ccccc cccc cccc |c|}
		\hline
		$k$ vs. $|\mathcal{O}|$  
		& 3  &  4 & 5  &  6 &7 &8 &9 &10 &11 &12 &13 &14 &15 &16 &17 & sum\\
		\hline
		2    & -  &  1 &2   &  2 &- &- &-  &- &-  &-  &-  &-  &-  &-  &-  & 5 \\		
		1.5  & -  &  1 &3   &  3 &- &3 &1  &- &-  &-  &-  &-  &-  &-  &-  & 11 \\
		1.2  & -  &  - &-   &  3 &8 &10&3  &- &-  &1  &-  &-  &-  &-  &-  & 25 \\
		1    & -  &  - &-   &  2 &6 &9 &7  &8 &9  &4  &-  &-  &-  &-  &-  & 45 \\
		0.9  & -  &  - &-   &  - &1 &6 &21 &18&11 &20 &-  &-  &-  &-  &-  & 77 \\
		0.8  & -  &  - &-   &  - &- &13&19 &37&37 &35 &24 &-  &-  &-  &-  & 165 \\
		0.7  & -  &  - &-   &  - &- &2 &17 &50&103&80 &71 &16 &3  &-  &-  & 342 \\
		0.6  & -  &  - &-   &  - &- &5 &2  &21&119&184&215&120&31 &-  &-  & 697 \\
		0.5  & -  &  - &-   &  - &- &- &-  &3 &34 &168&381&487&315&95 &5  & 1488 \\ 		\hline
	\end{tabular}}
\end{table}

We observe that for smaller scale, i.e. higher precision, the sizes of the stable order ideals increase.
Indeed, in the limit $k \rightarrow 0$, $S_{num}^{Fas}(D_Y,\delta)$ becomes the statistical fan $S(D_Y)$
and the sizes of the stable order ideals become $|D_Y|$. \\
We also compare the sizes of the numerical statistical fan $S_{num}^{Fas}(D_Y,\delta)$ to the  statistical fan $S(D_Y)$.
Table \ref{tab:compare} shows the size of the numerical statistical fan (for  $\delta=(12.5, 7,0.3, 1.5)$)
and for the  statistical fan whose cardinality grows exponentially with the size of the  design. Here, we successively deleted arbitrary points from the design with $|D_Y|=17$. Furthermore, we also computed the number of all hierarchical models with $n$ (and with $\le n$) terms, denoted by
$p_d(n)$ and $p_d(\le n)$, respectively.
For $d=4$ they correspond to so-called solid partitions. Recall that there is no generating function of $p_d(n)$ known for $d \ge 4$. First enumerations of the number of solid partitions were done in \cite{ABMM67} and \cite{Kn70}. 

\begin{table}[t!]
		\caption{Comparison of numerical and statistical fan sizes}
	\label{tab:compare}
\vskip .2cm
\centerline{\tabcolsep=3truept
	\begin{tabular}{|c|c|c||c|c|c|c|}
		\hline
		$|D_Y|$ & $|S_{num}^{Fas}(D_Y,\delta)|$ & \#stable OIs & $|S(D_Y)|$ &  \#OIs & $p_d(n)$ & $p_d(\le n)$\\
		\hline
		17 & 45 & 481   & 416570 & 847078 & 416849 & 847517 \\
		16 & 39 & 402   & 213954 & 430495 & 214071 & 430668 \\
		15 & 45 & 425   & 108752 & 216529 & 108802 & 216597 \\
		14 & 34 & 339   &  54791 & 107777 &  54804 & 107795 \\
		13 & 33 & 317   &  27235 &  52973 &  27248 &  52991 \\
		12 & 34 & 298   &  13413 &  25725 &  13426 &  25743 \\
		11 & 24 & 206   &   6487 &  12299 &   6500 &  12317 \\
		10 & 22 & 191   &   3109 &   5799 &   3122 &   5817 \\
		9  & 33 & 179   &   1451 &   2677 &   1464 &   2695 \\
		8  & 21 & 118   &    680 &   1226 &    684 &   1231 \\
		\hline
	\end{tabular}}
\end{table}
We observe that the numerical statistical fan is significantly smaller than the  statistical fan, i.e. it is
effectively computable.
Furthermore, its size does not grow (sub-)exponentially with the design size.
The size of the statistical fan of $D_Y$ turns out to be only a bit smaller than the number of all maximal hierarchical models $p_d(n)$. Similarly, the number of all order ideals identifiable by $D_Y$ is only a few less than  the number of all hierarchical models (with less than $n$ terms). This shows that $D_Y$ is close to a truly generic design where we would expect that 
	$|S(D_Y)|$ and $p_d(n)$ coincide.

\section{Discussion and Outlook}
\label{sec:OutlookAndDiscussion}

The computation of  the numerical fan $S_{num}(D,\delta)$ provides improved model selection through the enhanced knowledge of the space of all identifiable stable hierarchical models that avoid "numerical aliasing".
In stark contrast to the statistical fan of a generic design, the numerical fan turns out to be effectively
computable - at least for small data sets with few dimensions.
The recursive enumeration of the numerical fan $S_{num}(D,\delta)$ can be combined with subset selection by considering
only submodels that contain the new monomial $t$ to be included to an order ideal $\mathcal{O}$.
Maximal Models can also be roughly ordered according to their validity by scaling the tolerance vector. \\
In future work we may consider an extension to other model classes like quasi-order ideals \citep{MT12}.
The recursive algorithm can be used to search for low degree polynomials that describe varieties close to all empirical points. Indeed, every numerical linear dependence provides such polynomial equations. Recall that the Fassino condition 
(with $\delta ^2$-term) is only a sufficient condition for numerical independence. If it is not fullfilled, we may search for a pertubed design providing linear dependence of design vectors. Such a search algorithm was suggested in \cite{FT13}.

\section*{Acknowledgements}

The financial support of the Deutsche Forschungsgemeinschaft (SFB 823, project B1) is gratefully acknowledged.

\vskip .65cm
\noindent
Arkadius Kalka, Dortmund University of Applied Sciences and Arts.
\vskip 2pt
\noindent
E-mail: (arkadius.kalka@fh-dortmund.de)
\vskip 2pt

\noindent
Sonja Kuhnt, Dortmund University of Applied Sciences and Arts.
\vskip 2pt
\noindent
E-mail: (sonja.kuhnt@fh-dortmund.de)

\end{document}